\documentclass[onecolumn,usenatbib]{mn2e}
\usepackage{graphicx}
\usepackage[a4paper,twosideshift=0pt,totalwidth=480pt, totalheight=680pt]{geometry}

\begin{document}
\def\n{\mathrm{n}}
\def\p{\mathrm{p}}
\def\e{\mathrm{e}}
\def\b{\mathrm{b}}

\def\d{\delta}
\def\A{{\cal A}}
\def\B{{\cal B}}
\def\C{{\cal C}}
\def\y{\Lambda}
\def\pat{\sigma_p}

\def\mun{{\mu_\n}}
\def\mup{{\mu_\p}}

\def\Lsph{{\nabla^2_\theta}}
\def\Ulm{{U_l^m}}
\def\Vlm{{V_l^m}}
\def\Wlm{{W_l^m}}
\def\tVlm{{\tilde{V}_l^m}}
\def\tWlm{{\tilde{W}_l^m}}
\def\tUlm{{\tilde{U}_l^m}}

\def\dvnr{\delta v^r_\n}
\def\dvnt{{\delta v^\theta_\n}}
\def\dvnp{{\delta v^\phi_\n}}

\def\dvpr{\delta v^r_\p}
\def\dvpt{{\delta v^\theta_\p}}
\def\dvpp{{\delta v^\phi_\p}}

\def\sint{{\rm sin} \theta}
\def\cost{{\rm cos} \theta}
\def\cott{{\rm cot} \theta}
\def\tant{{\rm tan} \theta}

\def\parr{{\partial \over \partial r}}
\def\part{{\partial \over \partial \theta}}
\def\parp{{\partial \over \partial \phi}}

\def\E{{\mathcal{E}}}
\def\mut{\widetilde{\mu}}
\def\eps{\varepsilon}

\title{The superfluid two-stream instability and pulsar glitches}

\author[N. Andersson, G.L. Comer and R. Prix]
{N. Andersson$^\dag$, G.L. Comer$^*$ and R. Prix$^\dag$\\
$^\dag$Department of Mathematics, University of Southampton, Southampton, SO17 1BJ, United 
Kingdom \\$^*$Department of Physics, Saint 
Louis University, St Louis, MO 63156,  USA}

\maketitle

\begin{abstract}
This paper provides the first study of a new dynamical instability in
superfluids. This instability is similar to the two-stream instability
known to operate in plasmas. It is analogous to the Kelvin-Helmholtz
instability, but has the distinguishing feature that 
the two fluids are interpenetrating. The instability sets in once the relative 
flow between the two components of the system reaches a critical level. 
Our analysis is based on the two-fluid equations
that have been used to model the dynamics of the outer core 
of a neutron star, where superfluid neutrons are expected to coexist
with superconducting protons and relativistic electrons. These equations
are analogous to the standard Landau model for superfluid Helium.
We study this instability for two different model problems.  
First we analyze a local dispersion relation for waves in a system
where one fluid is at rest while the other flows at a constant rate.
This provides a proof of principle of the existence of the two-stream 
instability for superfluids. Our second model problem concerns two rotating 
fluids confined within an infinitesimally thin spherical shell. 
The aim of this model is to assess whether 
the two-stream instability may be relevant (perhaps as a trigger mechanism)
for pulsar glitches. Our results for this problem show 
that the entrainment effect could provide a sufficiently strong coupling 
for the instability
to set in at a relative flow small enough to be astrophysically 
plausible.
\end{abstract}

\maketitle

\section{Introduction}

In this paper we describe a new dynamical instability 
in superfluids.  This ``two-stream'' instability is analogous to the
Kelvin-Helmholtz instability \citep{drazin}. 
Its key distinguishing feature is that the two fluids 
are interpenetrating rather than in contact across an 
interface as in the standard scenario. 
The two-stream instability is well known in plasma 
physics [where it is sometimes referred to as the ``Farley-Buneman'' 
instability 
\citep{farley,buneman,buneman2}], and it has also been discussed in various 
astrophysical 
contexts like merging galaxies \citep{galax}
and pulsar magnetopheres \citep{pulsar1,pulsar2,pulsar3}, but as far as we 
are aware it has not
been previously considered for superfluids. 
In fact, the ``standard'' Kelvin-Helmholtz instability was only recently 
discussed in the context of superfluids \citep{kh}.

The similarity of the equations used in plasma physics [a nice pedagogical 
description of the plasma two-stream instability can be found in 
\cite{anderson}]
to the ones for two-fluid superfluid models inspired us to ask whether an 
analogous instability could be relevant for
 superfluids. That this ought to be the case seemed inevitable. 
To prove the veracity of this expectation, we 
have adapted the arguments from the plasma problem to the superfluid case, 
and discuss various aspects of the two-stream instability in this paper.

Of particular interest to us is the possibility that the two-stream 
instability may operate in rotating superfluid neutron stars.
Mature neutron stars are expected to be 
sufficiently cold (eg. below $10^9$~K) that their
interiors contain several superfluid/superconducting 
components. Such loosely coupled components are usually invoked
to explain the enigmatic pulsar glitches, sudden spin-up 
events where the observed spin rate jumps by as much as
one part in $10^6$ \citep{lyne}. Theoretical models for glitches 
\citep{ruder,bppr,ai}
have been discussed ever since the first Vela pulsar glitch was observed
in 1969 \citep{vela1,vela2}, but these event are still not well understood. 
After three decades of theoretical effort it is generally accepted that the 
glitches arise because a superfluid component can rotate 
at a rate different from that of the bulk  of the star, and that 
a transfer of angular momentum from the superfluid to the crust 
of the star could lead to the observed phenomenon. 
The relaxation
following the glitch is well explained in terms 
of vortex creep [see for example \citet{vcreep}], but the mechanism that 
triggers the glitch event remains elusive. In this context, it seems
plausible that  
the superfluid two-stream instability may turn out to be relevant. 

\section{Proof of principle: A local analysis}

\subsection{Superfluid dispersion relation}

We take as our starting point the two-fluid 
equations that  have been used to discuss the dynamics
of superfluid neutron stars \citep{mendell1,ac01,prix02}.
Hence, we consider superfluid neutrons (index $\n$) coexisting with 
a conglomerate of charged components (index $\p$). 
The constituents of the latter (mainly ions and electrons in 
the neutron star crust and protons and electrons in the core)
are expected to be coupled by viscosity and the magnetic
field on a very short timescale. Hence, we assume that these charged
components will move together and that it is appropriate to treat them
as a single fluid. 

The corresponding equations are \citep{ac01,prix02}
\begin{equation}
\partial_t n_X + \nabla\cdot (n_X \vec{v}_X) = 0\,,
\label{fullcons}
\end{equation}
where $n_X$ represent the respective number densities and $\vec{v}_X$ are the 
two velocities. Here, and in the following, we use the constituent index $X$
which can be either $\n$ or $\p$. The respective mass densities are
obviously given by \mbox{$\rho_X = m_X\, n_X$} and we further introduce
the relative velocity $\vec{\Delta}$ between the two fluids as 
\begin{equation}
\vec{\Delta} \equiv \vec{v}_\p - \vec{v}_\n \ .
\end{equation}
The first law of thermodynamics is defined by the differential of the
energy density or ``equation of state'', 
\mbox{$\E=\E(\rho_\n, \rho_\p, \Delta^2)$}, namely
\begin{equation}
  \label{eq:FirstLaw}
  d\E = \mut^\n\, d\rho_\n + \mut^\p \, d\rho_\p + \alpha\, d\Delta^2\,,
\end{equation}
which defines the chemical potentials $\mut^X$ and the ``entrainment''
 function $\alpha$ as the thermodynamic conjugates to the densities and
the  relative velocity.
 With these definitions we can write the two Euler-type equations:
\begin{eqnarray}
  (\partial_t + \vec{v}_\n\cdot\vec{\nabla}) \left(\vec{v}_\n + \eps_\n \vec{\Delta} \right) 
  + \vec{\nabla} \left(\Phi + \mut^\n \right) + \eps_\n
  \Delta_j \vec{\nabla} v_\n^j &=& 0\,,\\ 
  (\partial_t + \vec{v}_\p\cdot\vec{\nabla}) \left(\vec{v}_\p - \eps_\p \vec{\Delta} \right) 
  + \vec{\nabla} \left(\Phi + \mut^\p \right) - \eps_\p
  \Delta_j \vec{\nabla} v_\p^j &=& 0\,,
\end{eqnarray}
where we have introduced the dimensionless entrainment parameters
\begin{equation}
\eps_X \equiv {2 \alpha \over  \rho_X}\,.
\end{equation}
The equation for the gravitational potential $\Phi$ is
\begin{equation}
   \nabla^2 \Phi = 4 \pi G \rho  \ , \label{poisson}
\end{equation}
where $\rho = \rho_\n +\rho_\p$. 

When $\alpha \neq 0$ these equations make manifest the so-called entrainment effect.
The entrainment  arises 
because the bare neutrons (or protons) are "dressed" by a polarization cloud of nucleons 
comprised of both neutrons and protons.  Since both types of nucleon contribute to the 
cloud the momentum of the neutrons, say, is modified so that it is a linear combination 
of the neutron and proton particle number density currents (the same is true for the 
proton momentum).  This means that when one of the fluids begins to flow it 
will, through entrainment, induce a momentum in the other.
Because of entrainment a 
portion of the protons (and electrons) will be pulled along with the superfluid neutrons 
that surround the vortices by means of which the superfluid mimics 
bulk rotation. This motion leads to magnetic fields being attached to the 
vortices, and dissipative scattering of electrons off of 
these magnetic fields. This ``mutual friction'' is expected to 
provide one of the main dissipative mechanisms in superfluid neutron star cores \citep{mendell2}.

In order to establish the existence of the superfluid two-stream instability
we consider the following model problem: Let the unperturbed configuration be 
such that the ``protons'' remain at rest, while the neutrons flow with 
a constant velocity $v_0$. For simplicity, we neglect the
 coupling through entrainment , 
i.e. we take $\alpha =0$,  and we also neglect perturbations in the
gravitational potential. 
Under these assumptions, the two fluids are only coupled
``chemically'' through the equation of state.

Writing the two velocities as  
$\vec{v}_\n = [v_0+\delta v_\n(t,x)]\hat{x}$ and $\vec{v}_\p =\delta v_\p(t,x)\hat{x}$
where $\delta v_\n$ and $\delta v_\p$ are taken to be suitably small, 
we get the perturbation equations
\begin{eqnarray}
\partial_t \delta n_\n + v_0 \partial_x \delta n_\n + n_\n \partial_x \delta v_\n &=& 0 \ , \\
\partial_t \delta n_\p  + n_\p \partial_x \delta v_\p &=& 0 \ , 
\end{eqnarray}
and
\begin{eqnarray}
\partial_t \delta v_\n + v_0 \partial_x \delta v_\n + \partial_x  \delta \mut^\n  &=& 0 \ , \\
\partial_t \delta v_\p  + \partial_x  
\delta \mut^\p  &=& 0 \ .
\end{eqnarray}
Next, we  assume harmonic dependence on both 
$t$ and $x$, i.e. we use the Fourier decomposition
$\delta v_X(t,x) = \delta v_X \exp [i(\omega t - kx)]$ etcetera.
This leads to the four equations
\begin{eqnarray}
i(\omega - kv_0) \delta n_\n - ik n_\n \delta v_\n &=& 0 \ , \\ 
i\omega \delta n_\p - ik n_\p \delta v_\p &=& 0 \ , \\
i(\omega -kv_0) \delta v_\n - ik \delta \mut^\n &=& 0 \label{eul_n} \ , \\
i\omega  \delta v_\p - ik \delta \mut^\p &=& 0 \label{eul_p} \ . 
\end{eqnarray}

We thus have four equations relating the six unknown variables $\delta v_X$, 
$\delta n_X$ and $\delta \mut^X$. 
To close the system we need to provide an equation of state. 
Given an energy functional ${\cal E} = {\cal E} (n_\n,n_\p)$ we have
\begin{equation} 
m \delta \mut^\n = \left. \left( { \partial \mun \over \partial n_\n }
\right) \right|_{n_\p} \delta n_\n + 
\left. \left( { \partial \mun \over \partial n_\p }
\right) \right|_{n_\n} \delta n_\p =  { \partial^2 {\cal E}  \over \partial n_\n^2 } \delta n_\n +  { \partial^2 {\cal E}  \over \partial n_\p \partial n_\n }
\delta n_\p
\end{equation}
and similarly 
\begin{equation}
 m \delta \mut^\p =  { \partial^2 {\cal E}  \over \partial n_\p \partial n_\n }
\delta n_\n +  { \partial^2 {\cal E}  \over \partial n_\p^2 } \delta n_\p \ .
\end{equation}
Finally, we define the two sound speeds by, cf. \citet{ac01},
\begin{eqnarray}
c_\n^2  &=& n_\n \left. { \partial \mut^\n \over \partial n_\n}
\right|_{n_\p} = { n_\n \over m} { {\cal S}_{\p\p} \over \det {\cal S}} \ , \\
c_\p^2  &=& n_\p \left. { \partial \mut^\p \over \partial n_\p}
\right|_{n_\p}  = { n_\p\over m} { {\cal S}_{\n\n} \over \det {\cal S}} \ , 
\end{eqnarray}
and introduce the ``coupling parameter'' 
\begin{equation}
{\cal C}  = n_\n \left. { \partial \mut^\n \over \partial n_\p}
\right|_{n_\n} = n_\n \left. { \partial \mut^\p \over \partial n_\n}
\right|_{n_\p}= -  { n_\n \over m} { {\cal S}_{\n\p} \over \det {\cal S}}\\
\end{equation}
which also has the dimension of a velocity squared.
For later convenience we have given the relation to the coefficients
of the ``structure matrix'' ${\cal S}_{XY}$ used by  \citet{prix}.

With these definitions we get 
\begin{eqnarray}
n_\n \delta \mut^\n &=&  c_\n^2  \delta n_\n + {\cal C} \delta n_\p \ , \\
n_\p \delta  \mut^\p &=& { n_\p \over n_\n}  {\cal C}  \delta n_\n + c_{\p}^2 \delta n_\p \ ,
\end{eqnarray}
and we can rewrite our set of equations 
as
\begin{eqnarray}
n_\n \delta v_\n &=& \left({ \omega \over k} - v_0 \right) \delta n_\n = 
 \left({ \omega \over k} - v_0 \right)^{-1} \left[ c_\n^2  \delta n_\n + {\cal C} \delta n_\p \right] \ , \\
n_\p \delta v_\p &=& { \omega \over k} \delta n_\p = { k \over \omega} 
\left[  { n_\p \over n_\n}  {\cal C}  \delta n_\n + c_{\p}^2 \delta n_\p\right] 
\ . 
\end{eqnarray}
Reshuffling we get
\begin{eqnarray}
\left[ \left({ \omega \over k} - v_0 \right)^2 - c_\n^2 \right] \delta n_\n &=& 
 {\cal C} \delta n_\p \ , \\
\left[ \left( { \omega \over k}\right)^2 - c_\p^2 \right]\delta n_\p  &=&
  { n_\p \over n_\n}  {\cal C} \delta n_\n \ , 
\end{eqnarray}
and a dispersion relation
\begin{equation}
\left[ \left({ \omega \over k} - v_0 \right)^2 - c_\n^2 \right]\left[ \left( { \omega \over k}\right)^2 - c_\p^2 \right] =  { n_\p \over n_\n} {\cal C}^2 \ .
\end{equation}

Introducing the ``pattern speed'' (the phase velocity)
$\pat = \omega/k$ we have
\begin{equation}
\left[ (\pat - v_0 )^2 - c_\n^2 \right]
\left[ \pat^2 - c_\p^2 \right] =  { n_\p \over n_\n} {\cal C}^2 \ . 
\label{disp0}
\end{equation}
Not surprisingly, 
this local dispersion relation is qualitatively similar to the one for the 
plasma problem \citep{anderson}. We will now use it to investigate under what circumstances
we can have complex roots, i.e. a dynamical instability. 

\subsection{The superfluid two-stream instability}

First of all, it is easy to see that (\ref{disp0})  leads to 
the simple roots
\begin{equation}
\pat = \left\{ \begin{array}{ll} \pm c_\p \\ v_0 \pm c_\n \end{array}
\right.
\end{equation}
in the uncoupled case, when ${\cal C} = 0$.
This establishes the interpretation of $c_X$ as the sound speeds. 

To investigate the coupled case, we introduce new variables
$x=\pat/c_\n$ and $y=v_0/c_\n$. Then we get 
\begin{equation}
f(x,y) = { 1 \over a^2} [ (x-y)^2-1][x^2-b^2] = 1
\label{fxy}\end{equation}
where we have defined
\begin{equation}
a^2 \equiv { n_\p \over n_\n} {{\cal C}^2 \over c_\n^4}
\quad  \mbox{ and } \quad b^2 \equiv { c_{\p}^2 \over c_\n^2} \ .
\label{adef}
\end{equation}

The onset of dynamical 
instability typically corresponds to the merger of two 
real-frequency modes. If this is the case, a marginally stable configuration
will be such that  (\ref{fxy}) has a double root. This happens when 
an inflexion point of  $f(x,y)$ coincides with $f(x,y)=1$.
This is a useful criterion for searching for the marginally stable 
modes of our system.

\begin{figure} 
\centering
\includegraphics[height=6cm,clip]{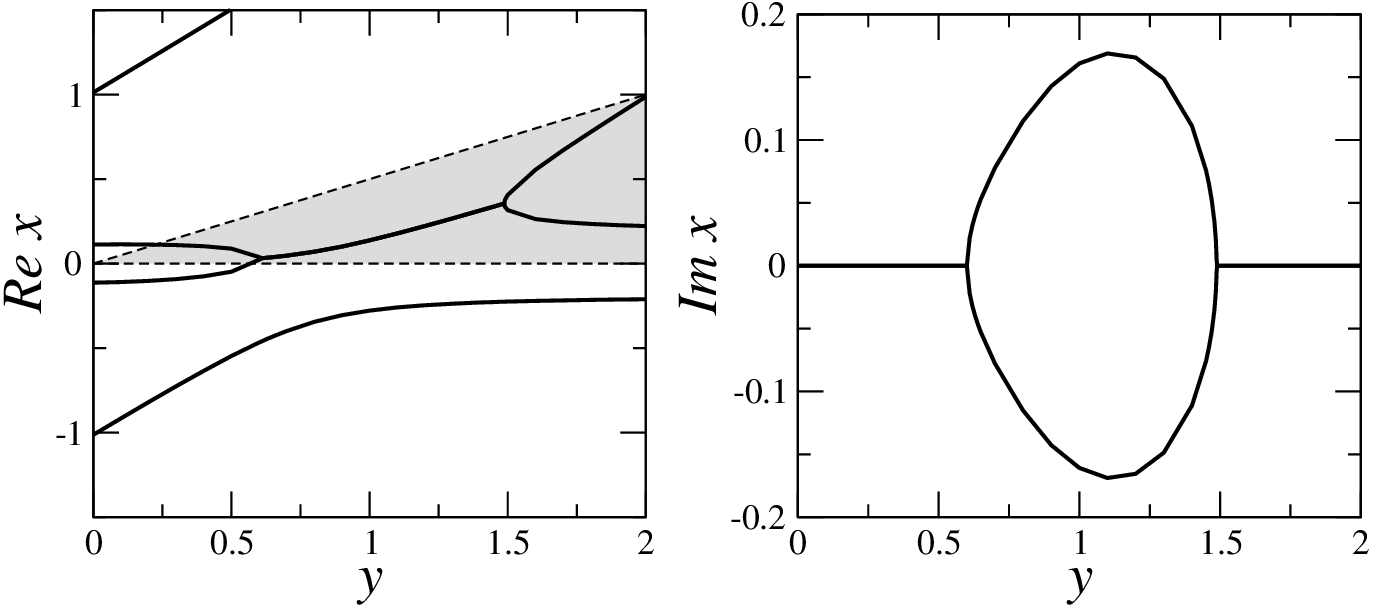}
\caption{Real (left panel) and imaginary (right panel) parts of the four roots of the 
dispersion relation (\ref{fxy}) for  model parameters
 $a^2=0.0249$ and $b^2=0.0379$. For these parameters the quartic dispersion relation 
has four real roots for both
$y=0$ and $y=2$, while it has two real roots and a complex
conjugate pair for $y$ in the range $0.6<y<1.5$. In this range, the 
two-stream instability is operating. The grey area corresponds to
$0 \le \mbox{Re } x \le 1/2$ which is contained in the range of the 
instability criteria discussed in Section~IIC. }
\label{modes}
\end{figure}

As a ``proof of principle'' we consider the particular case
of  $a^2=0.0249$ and $b^2=0.0379$ (we will motivate this particular choice 
in Section~IIE). 
The real and imaginary parts of the mode-frequencies for these 
parameter values are shown as functions 
of $y$ in Figure~\ref{modes}. We have complex roots (an instability) 
in the range $0.6 < y < 1.5$.
The corresponding mode frequencies lie in the range $0.03 < x_0< 0.36$. The 
fastest growing instability occurs for $y \approx 1.1$ for which
we find that $\mbox{Im } x \approx 0.17$. In other words, in this particular
case we encounter the two-stream instability once the  rate of the background flow is increased 
beyond
\begin{equation}
v_0 = c_\n y \approx 0.6 c_\n \ . 
\end{equation}
The corresponding frequency is given by
\begin{equation}
\omega = k c_\n x_0 \approx 0.1 k c_\n \ .  
\end{equation}
From this we see that the instability is
present  well before the neutron flow becomes ``supersonic''. This is crucial
since one would expect the superfluidity to be destroyed for 
supersonic flows. 

We have thus 
established that the two-stream instability may, in principle, operate in 
superfluids. Our example indicates the existence of a lower limit 
of the background flow for the instability. This turns out to be 
a generic feature.  
In contrast, the plasma two-stream instability  can operate
at arbitrarily slow flows. An ideal plasma is unstable to 
sufficiently long wavelengths for any given $v_0$. 
In reality, however, one must also account for dissipative mechanisms.
In the case of real plasmas one finds that the so-called Landau 
damping stabilizes 
the longest wavelengths \citep{anderson}. Thus the two-stream instability
sets in below a critical wavelength in more realistic plasma models, 
and there is typically (just like in the present case)
a range of flows for which the instability is present.
We will discuss the effects of dissipation on the superfluid two-stream 
instability briefly in Section~IV.

\subsection{Necessary criteria for instability}

It is useful to consider whether we can derive a necessary condition 
for the two-stream instability.
To approach this problem in full generality would likely be quite
complicated, but we can  make good progress for the simple
one-dimensional toy problem discussed above. 

We begin by multiplying the Euler equation (\ref{eul_n}) for the neutrons by the complex conjugate  
$\delta v^*_\n$. This leads to (after also using the continuity equations to replace
the perturbed number densities) 
\begin{equation}
\left( \omega -k v_0 - { k^2 c_\n^2 \over \omega -kv_0} \right) |\delta v_\n|^2 = 
{\cal C} {k^2 n_\p \over n_\n \omega} \delta v^*_\n \delta v_\p \ . 
\end{equation}
Similarly, we obtain from the second Euler equation (\ref{eul_p})
\begin{equation}
\left( \omega  - { k^2 c_\p^2 \over \omega} \right) |\delta v_\p|^2 = 
{\cal C} {k^2 \over\omega - kv_0} \delta v^*_\p \delta v_\n \ . 
\end{equation}
Next we combine these two equations to get
\begin{equation}
{\cal L} = { n_\n \over n_\p} \pat \left( \pat - v_0 - { c_\n^2 \over \pat-v_0} \right)
| \delta v_\n|^2 + (\pat-v_0) \left( \pat - { c_\p^2 \over \pat} \right) | \delta v_\p|^2 = 
{\cal C} \left( \delta v^*_\n \delta v_\p + \delta v^*_\p \delta v_\n \right)  
\end{equation}
where we have introduced the pattern speed $\pat$. From this expression we 
see that the imaginary part of the left-hand side must vanish, so we should 
have Im~${\cal L}=0$.
Allowing the pattern speed to be complex, i.e. using $\pat = \sigma_R + i \sigma_I$ we find that 
the following condition must be satisfied
\begin{equation}
\mbox{Im } {\cal L} = \sigma_R \sigma_I \left\{ { n_\n \over n_\p}
  \left[ 2 - { v_0 \over \sigma_R} \left(1- { c_\n^2 \over
  |\pat-v_0|^2} \right)\right] | \delta v_\n|^2 + \left[ 2 - { v_0
  \over \sigma_R} \left( 1 + { c_\p^2 \over |\pat|^2} \right)\right]  
| \delta v_\p|^2  \right\} = 0 \ . 
\end{equation}
If we are to have an unstable mode, $\sigma_I\neq 0$, the frequency clearly 
must be such that  the factors multiplying the absolute values of the two velocities
have different signs.  

Let us first consider the case when the 
factor multiplying $|\delta v_\n|^2$ is negative. Then we find that an instability is
only possible if ${\sigma_R/ v_0} <0$, and the following
condition is satisfied:
\begin{equation}
0 < \left| {\sigma_R \over v_0} \right| < {1 \over 2} \left( 
{ c_\n^2 \over |\pat -v_0|^2} - 1 \right) \ . 
\end{equation}
This shows that we must have
\begin{equation}
{ c_\n \over |v_0|} > \left| { \pat \over v_0} -1 \right| >
\left| {\sigma_R \over v_0} - 1 \right| = \left| {\sigma_R \over v_0} \right| + 1 > 1
\end{equation}
which constrains the permissible frequencies to the range $|\sigma_R|< c_\n-|v_0|$. 
Thus we see that the flow must be subsonic, i.e. $|v_0|<c_\n$.

In the case when the 
factor multiplying $|\delta v_\p|^2$ is negative we can 
only have an instability if ${\sigma_R/ v_0} >0$.  We also require
\begin{equation} 
0 <  {\sigma_R\over v_0} < {1 \over 2}  \left( 1 + { c_\p^2 \over |\pat|^2} \right) < 
{1 \over 2}  \left( 1 + { c_\p^2 \over \sigma_R^2} \right) 
\quad \mbox{ if } \quad 
|\pat-v_0|^2 < c_\n^2
\label{cond1}\end{equation}
or
\begin{equation}
{ 1 \over 2} \left( 1 - { c_\n^2 \over |\sigma_R -v_0|^2}\right)
<
{ 1 \over 2} \left( 1 - { c_\n^2 \over |\pat -v_0|^2}\right) <  
{\sigma_R\over v_0}  
< {1 \over 2}  \left( 1 + { c_\p^2 \over \sigma_R^2} \right) 
\quad \mbox{ if } \quad 
|\pat-v_0|^2 > c_\n^2 \ .
\end{equation}

\begin{figure} 
\centering
\includegraphics[height=6cm,clip]{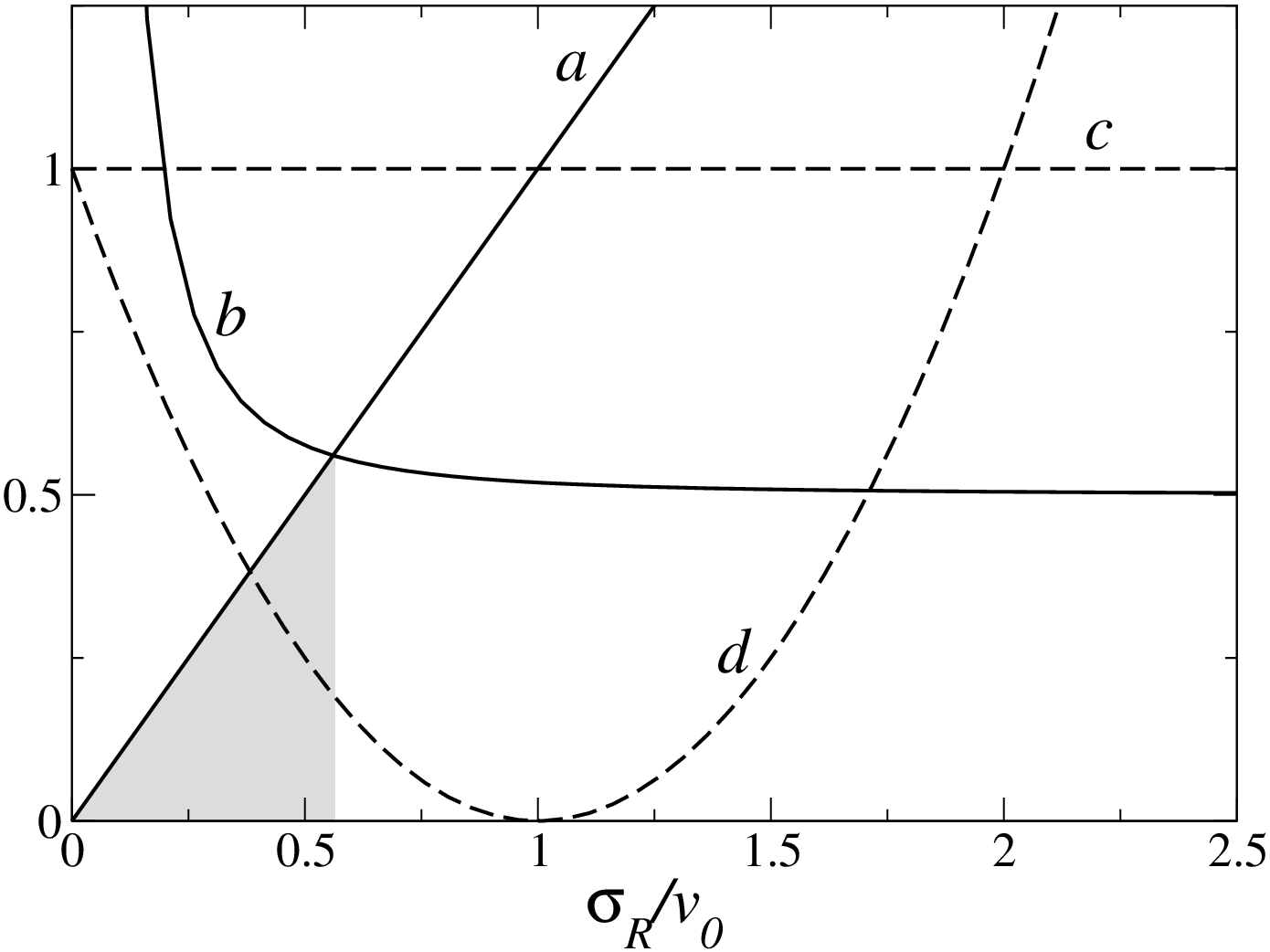}
\caption{An illustration of the instability criterion (\ref{cond1}) 
which is relevant for the example considered in Section~IIB.
This example is constructed by introducing $z=\sigma_R/v_0$, and then showing
the four curves: $a(z)=z$, $b(z)=(1+\gamma^2/z^2)/2$, $c(z)= 1$ and 
$d(z)=(z-1)^2$. 
Here we have taken $c_\n^2/v_0^2=1$ and $\gamma^2= c_\p^2/c_\n^2=0.0379$.
Criterion (\ref{cond1}) is satisfied when $d<c$ and $a<b$ (in the grey area). 
The corresponding range is well approximated by  $0<\sigma_R/v_0<1/2$.  }
\label{crit}
\end{figure}

For the example illustrated in Figure~\ref{modes}, the condition that must be 
satisfied is (\ref{cond1}). It is useful to notice two things about this criterion. 
First of all, any unstable mode for which $\sigma_R>c_\p$ must lie in the range
$0<c_\p<\sigma_R<v_0$. Secondly, when $\sigma_R>>c_\p$ the permissible 
range will be well approximated by $0<\sigma_R<v_0/2$, cf. Figure~\ref{crit}. 
As is clear from Figure~\ref{modes} the unstable modes satisfy this last, and most severe,
criterion.

It is worth noting that the instability can be discussed in terms
of a simple energy argument [see \citet{casti} and \citet{pierce} for similar discussions in 
other contexts].  
After averaging over several wavelengths, the kinetic energy of the protons
is
\begin{equation}
E_\p \approx {m_\p n_\p \delta v_\p^2 \over 2} > 0 \ . 
\end{equation} 
Meanwhile we get for the neutrons
\begin{equation}
E_\n \approx { m_\n \over 2} (n_\n+\delta n_\n)(v_0+\delta v_\n)^2
\end{equation}
which leads to 
\begin{equation}
E_\n \approx 
{m_\n n_\n \over 2}\left[ v_0^2 + {\omega +kv_0 \over \omega-kv_0} \delta v_\n^2 \right]
\end{equation}
from which we see that the energy in the perturbed flow is smaller than the
energy in the unperturbed case, which means that we can associate the wave 
with a ``negative energy'', when
\begin{equation}
-v_0 < { \omega \over k} < v_0\ , \quad \mbox{i.e. } \quad -v_0 < \sigma_R < 
v_0 \ .  
\end{equation}
A wave that satisfies $0<\sigma_R<v_0$ moves forwards with respect to the protons 
but backwards according to an observer riding along with the 
unperturbed neutron flow. As we have seen above, the unstable modes in our problem
satisfy this criterion and hence it is easy
to explain the physical 
conditions required for the two-stream instability to be present.

\subsection{Results for a simple model equation of state}

Having established that the two-stream instability may be present in
superfluids, we want to assess to what extent one should expect this 
mechanism to play a role for astrophysical neutron stars. 
To do this we will consider two particular equations of state. 
The results we obtain illustrate different facets of what we expect to 
be a rich problem. 

We begin by making contact with our recent analysis of rotating superfluid models
 \citep{prix} as well as the study of oscillating
non-rotating stars by \citet{pr02}.
From the definitions above we have
\begin{eqnarray}
a^2 &=& { n_\p \over n_\n} { {\cal S}_{\n\p}^2 \over{\cal S}_{\p\p}^2} \ , \\
b^2 &=&  { n_\p \over n_\n} { {\cal S}_{\n\n} \over{\cal S}_{\p\p}} \ .  
\end{eqnarray}
We combine these results with the explicit structure matrix 
given in eq. (144) of \citet{prix}, which is based on a simple
``analytic'' equation of state. This leads to
\begin{eqnarray}
a^2 &=&   { n_\p \over n_\n} \sigma^2 \ , \\
b^2 &=& 1 + { \sigma (1-2x_\p) \over 1-x_\p} \ , 
\end{eqnarray}
where $x_\p=n_\p/(n_\p+n_\n)$ is the proton fraction, and
 $\sigma$ is defined by
\begin{equation}
\sigma = { {\cal S}_{\n\p} \over {\cal S}_{\p\p} } \ . 
\end{equation}
As discussed by \citet{prix}, 
$\sigma$ is related to the ``symmetry energy'' of the 
equation of state, cf. \citet{PAL88}.

\begin{figure} 
\centering
\includegraphics[height=6cm,clip]{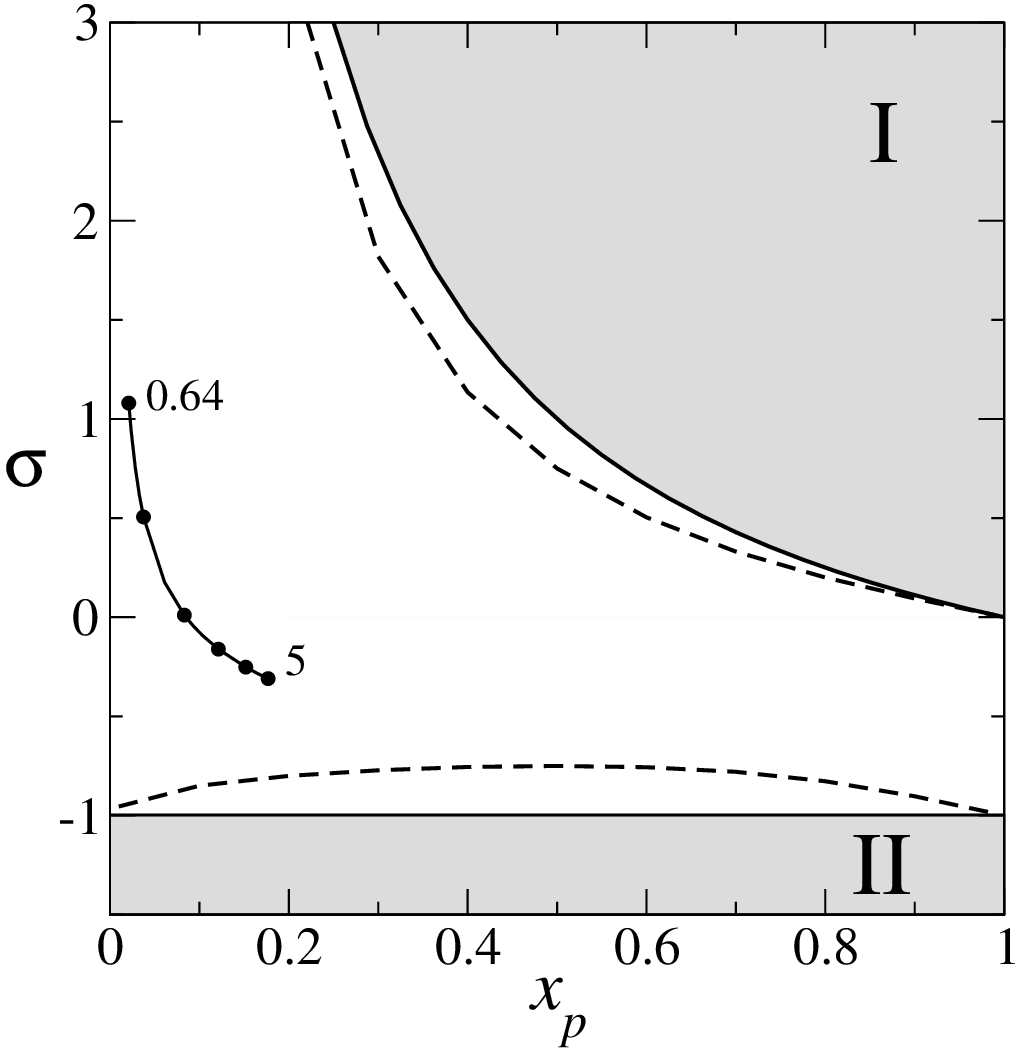}
\caption{An illustration of the various domains of instability 
for the simple ``analytic'' equation of state of \citet{prix}.
An absolute instability (see discussion in the main text) is active 
in the grey areas (also labeled I and II). The two-stream instability 
is, in principle, relevant in the remaining parameter space. 
The dashed curves indicate the onset of instability
when the relative flow is equal to the neutron sound speed $(y=1$). 
For slower flows, these critical curves approach the absolute instability
regions. The region where the two-stream instability may operate in 
physical flows therefore lies between each dashed curve and the nearest grey area.
For comparison we also indicate the curve in the $x_\p-\sigma$ plane traced out by the PAL 
equation of state (discussed in Section~IIE) 
as the density is varied from that near the crust/core interface
($u=0.64$) to five times that of nuclear saturation $(u=5)$.}
\label{domain}
\end{figure}

The instability regions for this model equation of state 
are illustrated in Figure~\ref{domain}. A key feature of this 
figure is the presence of regions of
 ``absolute instability''. This happens when $a^2>b^2$. Then 
there exist unstable solutions already for vanishing background 
flow, $y=0$.
That this is the case is easy to see. Consider $(\ref{fxy})$ in the limit
$y=0$. In the limit we can solve directly for $x^2$:
\begin{equation}
x^2 = { 1 +b^2 \over 2} \pm \sqrt{ \left( { 1 + b^2 \over 2} \right)^2 - b^2 + a^2  }  
\end{equation}
from which it is easy to see that one of the roots for  $x^2$
will be negative if $a^2>b^2$. Hence, one of the roots to the quartic
(\ref{fxy}) will be purely imaginary.

The physics of this instability is quite different from the
two-stream instability that is the main focus of this paper. 
Yet it is an interesting phenomenon. From the above relations 
we find that  $a^2 < b^2$ corresponds to 
\begin{equation}
{\cal S}_{\n\n}{\cal S}_{\p\p} > {\cal S}_{\n\p}^2 \ . 
\label{reason}
\end{equation}
In the discussion by \citet{prix} it was assumed that ``reasonable'' equations 
of state ought to satisfy this condition. We expected this to be the case 
since the structure matrix 
would not be invertible if its 
determinant were to vanish at some point. We now see that 
this constraint has a strong physical motivation: The condition is 
violated when $a^2>b^2$, i.e. when we have an absolute instability.
The regions where this instability is active are indicated by the grey areas
in Figure~\ref{domain}. 

\subsection{Results for the PAL equation of state}

In order to strengthen the argument that the two-stream instability
may operate in astrophysical neutron stars, we have considered a
``realistic'' equation of state due to  Prakash, Ainsworth and 
Lattimer (PAL) (1988). The advantage of this model is
that it is relatively simple. In particular, it leads to 
analytical expressions for the various quantities
needed in our analysis. The energy density of the baryons for the PAL
equation of state can be written
\begin{equation}
    {\cal E}(n_\n,n_\p) = \left(n_\n + n_\p\right) \left[E_0(u) + S(u) (1 - 
                      2 x_\p)^2\right] \ ,
\end{equation}
where $E_0$ corresponds to the energy per nucleon, $S$  
corresponds  to the ``symmetry energy'' (and is closely related to  $\sigma$ in the 
``analytic'' equation of state discussed above), and $u = (n_\n + n_\p)/n_0$ with  $n_0 = 0.16 
\ {\rm fm}^{-3}$ the nuclear saturation density.  
$E_0$ takes the following form: 
\begin{equation}
    E_0(u) = A_0 u^{2/3} + B_0 u + C_0 u^\sigma + 3 \sum_{i = 1}^2 C_i 
                \alpha_i^{-3} \left[\alpha_i u^{1/3} - \arctan 
                \left(\alpha_i u^{1/3}\right)\right] \ , 
\end{equation}
with $A_0 = 22.11$ MeV, $B_0 = 220.47$ MeV, $C_0 = - 213.41$ MeV, $\sigma 
= 0.927$, $C_1 = - 83.84$ MeV, $C_2 = 23.0$ MeV, $\alpha_1 = 2/3$, and 
$\alpha_2=1/3$.  The symmetry term is  
\begin{equation}
    S(u) = A_S \left[u^{2/3} - F(u)\right] + S_0 F(u) \ , 
\end{equation}
with $A_S = 12.99$ MeV and $S_0 = 30$ MeV. Here $F(u)$ is a 
function satisfying $F(1) = 1$ which is supposed to simulate the behaviour 
of the potentials used in theoretical calculations.  In our study 
we have only considered $F(u)=u$, which is one of four possibilities 
discussed by \citet{PAL88}.
 
We further need to account for the energy contribution of the
electrons, which is important since the electrons are 
highly relativistic inside neutron stars. Hence,  they can obtain 
 high (local) Fermi energies which may be comparable with the proton (local) 
Fermi energies.  Considering only the electrons, the leptonic 
contribution to the energy density is given by (in units where the speed 
of light is unity) \citep{shapiro}
\begin{equation}
    {\cal E}_{\e} = {m_\e \over \lambda_\e^3} 
                      \chi\left(\chi^{F}_\e\right) \ , 
\end{equation}
where $m_\e = m_\b/1836$ is the electron mass (in terms of the baryon mass 
$m_\b$), $\lambda_\e = \hbar/m_\e$ is the electron Compton wavelength, and 
\begin{equation}
    \chi(x) = {1 \over 8 \pi^2} \left(x \left[1 + x^2\right]^{1/2} 
              \left[1 + 2 x^2\right] - {\rm ln}\left[x + \left(1 + x^2
              \right)^{1/2}\right]\right) \ ,
\end{equation}
\begin{equation}
    \chi^{F}_\e = 1836 \left(3 \pi^2 \left[{\hbar \over m_b}\right]^3
                   \right)^{1/3} n_\p^{1/3} \ . 
\end{equation}
In doing this calculation  we have assumed local
charge neutrality, i.e.  $n_\e=n_\p$.
The above energy term is added linearly in the equation of state.

\begin{figure} 
\centering
\includegraphics[height=6cm,clip]{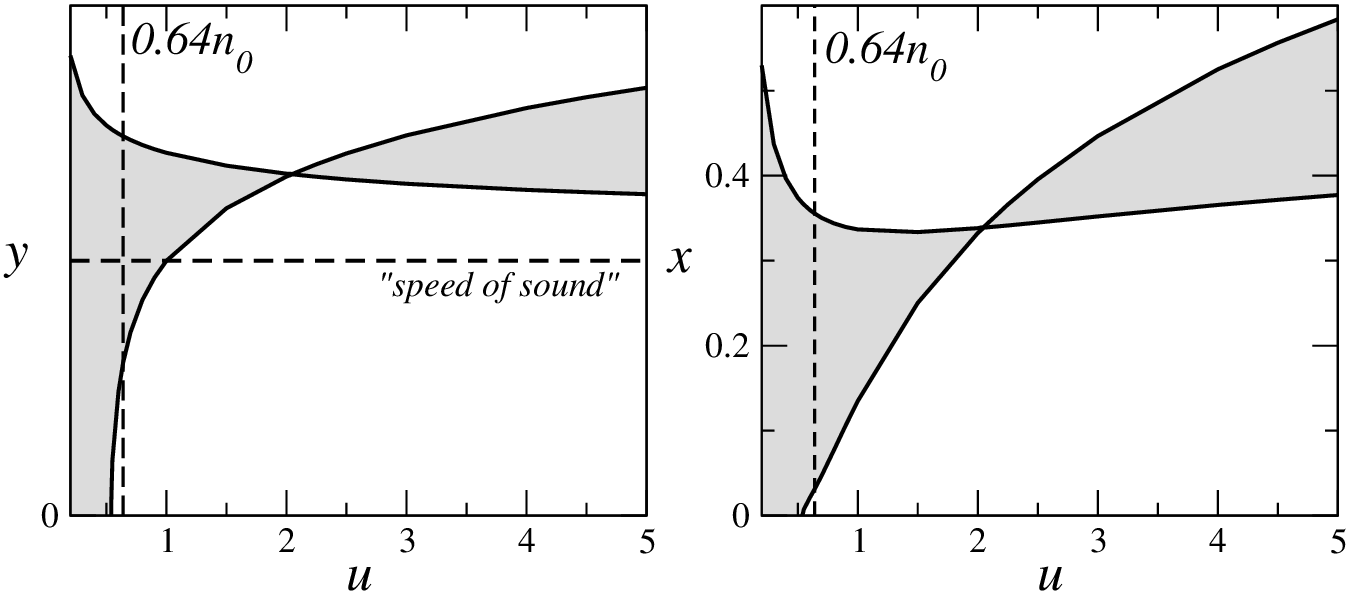}
\caption{Two-stream instability results for the PAL equation of state. 
Left panel: The region where the two-stream instability is present (grey area)
is shown as a function of the density parameter $u$. We indicate the location of the 
core-crust boundary ($u\approx 0.64$) by a vertical dashed line. Our model is only relevant 
for the core fluid, i.e. to the right of the vertical line. Finally, the horizontal dashed line indicates
when the relative flow is equal to the (neutron) sound speed. We expect the superfluid 
degeneracy to be broken beyond this level, so an instability located above this 
line is unlikely to have physical relevance. The results indicate that there
may be a region of instability immediately below the crust. Right panel: The corresponding 
oscillation frequencies. Particularly notable is the point near $u=2$ where the 
two critical curves cross. At this point the symmetry energy $\sigma$ changes sign, cf. 
Fig.~\ref{domain}, and there
exist a particular density such that the two fluids are uncoupled, cf. (\ref{adef}).  }
\label{local}
\end{figure}

Having obtained an expression for the total energy as a function of the 
density, we can derive explicit expressions for all quantities needed to 
discuss the two-stream instability. First we need to determine the proton fraction $x_\p$. We
do this by assuming that the star is in chemical equilibrium, i.e. 
\begin{equation}
\mu^\n = \mu^\p + \mu^\e\,.
\label{chemeq}\end{equation}
Solving (\ref{chemeq}) for $x_\p$ provides us with the proton fraction as
a function of $u$. Given this, and the relevant partial 
derivatives of ${\cal E+E}_\e$ we can readily evaluate the symmetry energy
and well as the sound speeds $c_\n^2$, $c_\p^2$ and the chemical 
coupling parameter ${\cal C}$. 
With this data we can determine the two parameters $a^2$ and $b^2$
which are needed if we want to solve the local dispersion relation (\ref{fxy}).
The results we obtain for the proton fraction and the symmetry energy are 
indicated in Figure~\ref{domain}. We consider the range from $u=0.64$, presumed to
correspond to the core-crust boundary, to $u=5$
which represents the deep core of a realistic neutron star.
The corresponding results for the two-stream instability are shown in Figure~\ref{local}.
From this figure we can see that the two-stream instability may operate 
(albeit at comparatively large relative flows) in the region immediately below
the crust. Finally,  we find that the conditions at the core-crust interface are such 
that $a^2=0.0249$ and $b^2=0.0379$. These are  the values we chose for the 
example in Section~IIB and hence the results shown in Figure~\ref{modes} 
correspond to a physically realistic model.

\section{Rotating shells: A connection with pulsar glitches?}

The fact that a superfluid neutron star may, in principle, exhibit the
two-stream instability does not necessarily prove that this 
mechanism will be astrophysically relevant.
Yet, it is an intriguing possibility given that
the mechanism underlying the enigmatic glitches observed in dozens of 
pulsars remains poorly understood. One plausible astrophysical 
role for the superfluid two-stream instability would be in this
context: Perhaps this instability serves as trigger 
mechanism for large pulsar glitches?

\subsection{Dispersion relation for rotating shells}

Our aim in this Section is to construct a toy problem that allows
us to investigate a possible connection between the 
two-stream instability and pulsar glitches. 
A suitably simple problem corresponds to two fluids,  allowed to rotate at different 
rates, confined within an infinitesimally thin spherical shell.
By assuming that the shell is infinitesimal we ignore radial motion, i.e. we restrict
the permissible perturbations of this system is such a way that 
the perturbed velocities must take the form 
\begin{equation}
    \delta \vec{v}_X 
=  - {1 \over R \sin \theta}  U^X_{lm}(t) 
                          \partial_\varphi Y^m_l \hat{e}_\theta 
+ {1 \over R }  U^X_{lm}(t) 
                        \partial_\theta Y^m_l  \hat{e}_\varphi
\end{equation} 
where  $Y^m_l(\theta,\varphi)$ are the spherical harmonics and $R$ is the radius $R$
of the shell. This means that the system permits only
toroidal mode-solutions. In other words, all oscillation modes
of this shell model are closely related to the inertial r-modes
of rotating single fluid objects \citep{pandp,provost}.

The perturbation equations for the configuration we consider 
have been derived in a different context and the complete 
calculation will be presented elsewhere \citep{inert2}. Our primary interest 
here concerns whether the modes of this system 
may suffer the two-stream instability. 
The presence of the instability in this
toy problem would be a strong indication that it will also be 
relevant when the shells are ``thick''
and radial motion is possible. 
That is, when we consider a rotating star 
that contains a partially decoupled superfluid either 
in the inner crust or the core. 

As discussed in Section~IIC, the two-stream instability 
can be understood in terms of negative energy waves. 
In the current problem, the criterion for waves to carry 
negative energy according to one fluid but positive 
energy according to the other fluid is that the mode pattern 
speed [we are assuming a decomposition $\exp(i\omega t + im\varphi)$]
\begin{equation}
\pat = - { \omega \over m}
\end{equation}
lies between the two (uniform) rotation rates. 
In other words, a necessary condition for instability is 
\begin{equation}  
\Omega_\p < \pat < \Omega_\n
\label{nec_crit}\end{equation}
where we have assumed that the superfluid neutrons lag behind the 
charged component, as is expected in a spinning down pulsar.
 
After a somewhat laborious calculation, 
see \citet{inert2} for 
details,  one finds that
the dispersion relation for the toroidal two-fluid modes of the 
shell problem is 
\begin{eqnarray}
    && \left\{ - l \left(l + 1\right) (1 - \eps_\n
          ) \left[\omega + m \Omega_\n\right] + 2 m \Omega_\n + m 
          \eps_\n \left[2 - l \left(l + 1\right)\right] 
          \left[\Omega_\p - \Omega_\n\right]\right\}\cr
       && \times \left\{ - l \left(l + 1\right) (1 - \eps_\p )
\left[\omega + m \Omega_\p\right] + 2 m 
          \Omega_\p + m \eps_\p \left[2 - l \left(l + 
          1\right)\right]\left[\Omega_\n - \Omega_\p\right]\right\} \cr
       && - \left\{ l \left(l + 1\right)\right)^2 \eps_\n \eps_\p 
\left(\omega + m \Omega_\n\right) \left(\omega + m 
          \Omega_\p\right\} = 0 \ .
\end{eqnarray}
This equation should be valid under the conditions in 
the outer core of a mature 
neutron star, where superfluid neutrons are
permeated by superconducting protons. 

We rewrite this dispersion relation in terms of the entrainment
parameter used by \citet{prix}, namely  
\begin{equation}
\eps = \eps_\p = { 2 \alpha \over \rho_\p} \ ,  
\end{equation}
the frequency as measured with respect to the rotation of the protons, 
\begin{equation}
\kappa = { \omega \over m\Omega_\p} \ , 
\end{equation}
and a dimensionless measure of the relative rotation, 
\begin{equation}
\y={ \Omega_\n \over \Omega_\p} \ . 
\end{equation} 
With these definitions we get
\begin{eqnarray}
&& \left\{ l(l+1) [1-x_\p(1+\eps)](\kappa+\y) -2(1-x_\p)\y + x_\p \eps(l-1)(l+2)
(1-\y) \right\} \nonumber\\
&& \times
 \left\{ l(l+1) [1-\eps](\kappa+1) -2 -  \eps(l-1)(l+2)
(1-\y) \right\} \nonumber \\
&& - [l(l+1)]^2 x_\p \eps^2 (\kappa+\y)(\kappa+1) = 0 \ . 
\label{disper}
\end{eqnarray} 
In terms of these new variables a mode would satisfy the necessary
instability criterion (\ref{nec_crit}) if $\kappa$ is such that 
\begin{equation}
-\y <\kappa< -1 \ .
\label{yrange}
\end{equation}
As we will now establish, there exist modes that satisfy this 
criterion for reasonable parameter values.

\subsection{An extreme example: The quadrupole modes}

\begin{figure} 
\centering
\includegraphics[height=6cm,clip]{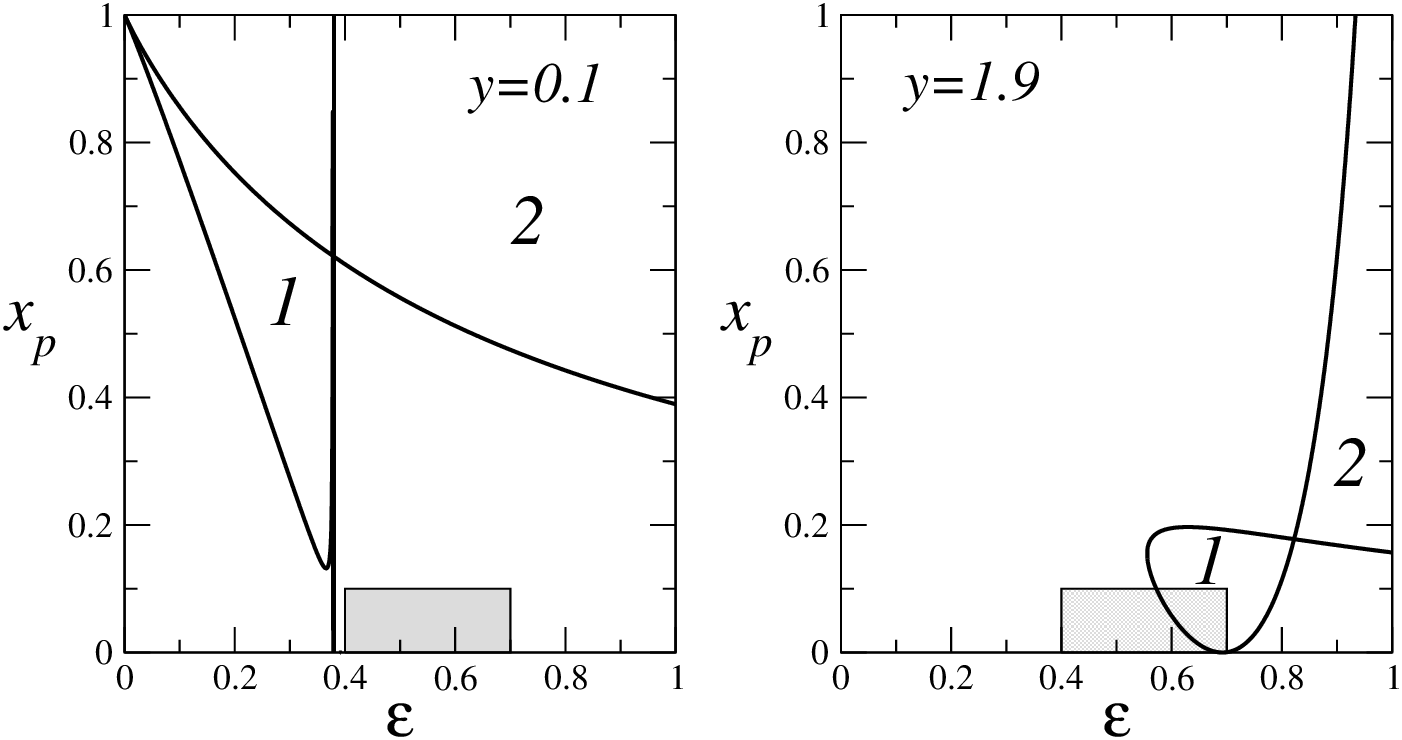}
\caption{Instability regions for the shell problem. We show 
results for $l=2$ and relative rotation such that 
the neutrons rotate 90\% slower (left frame) or faster (right frame) 
than the protons. The two-stream instability operates in 
regions 1-2. The grey box corresponds to the ``physically reasonable''
part of parameter space (for the core of a neutron star).}
\label{shell}
\end{figure}

\begin{figure} 
\centering
\includegraphics[height=6cm,clip]{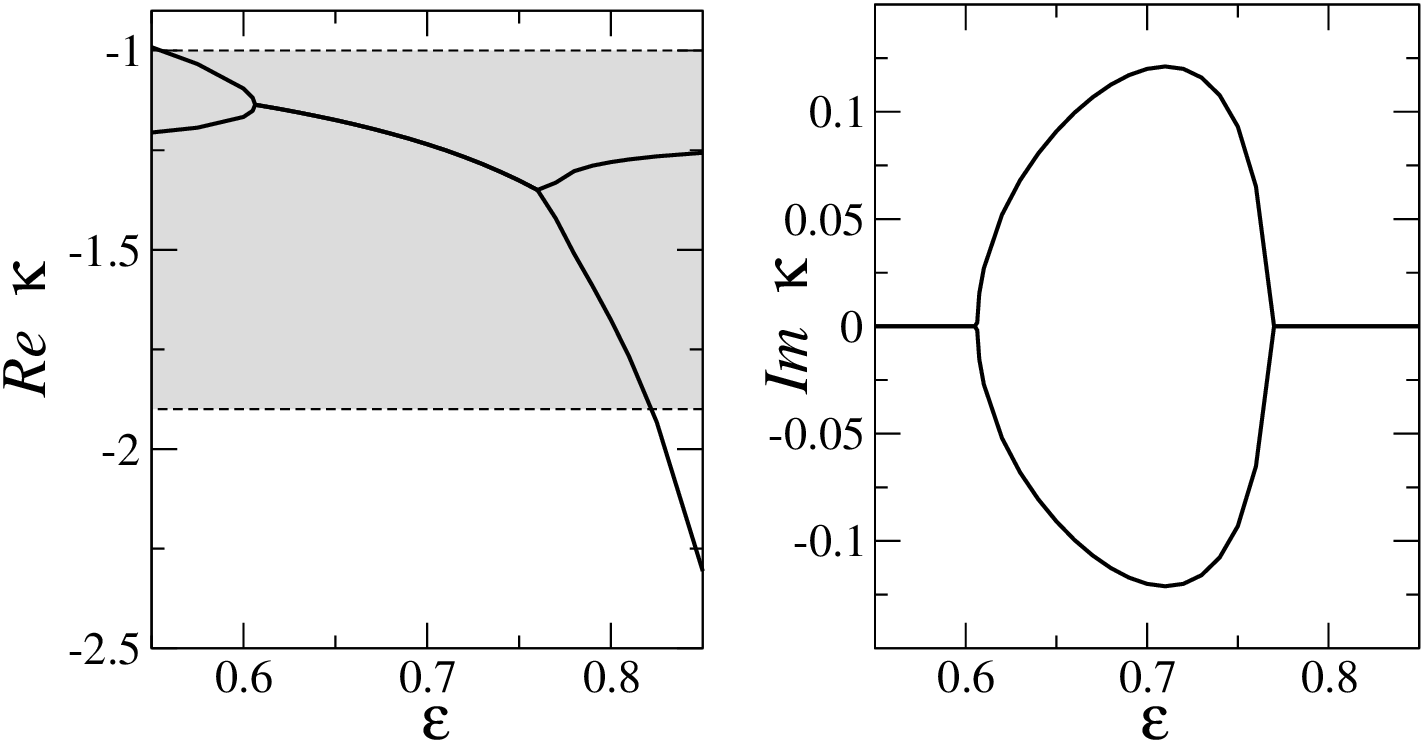}
\caption{Real and imaginary parts (left and right frame, respectively)
of the mode frequencies for $l=2$, $\y=1.9$ and $x_\p=0.05$. The
  corresponding instability region can be deduced from the 
right panel  of Fig.~\ref{shell}. The grey area in the left frame indicates the region
where an instability is permissible according to (\ref{yrange}). 
}
\label{rotfreq}
\end{figure}

Let us first consider the case of quadrupole oscillations, i.e. take $l=2$. 
Typical results for this case are shown in Figures~\ref{shell} and  \ref{rotfreq}.
The first figure illustrates the regions of the $x_\p-\eps$ parameter
space for which an instability is present in the case when (i) 
the neutrons rotate at a rate that is 90\% faster than that of the 
charged component, and (ii) the neutrons lag behind by the same fraction.
The second figure shows the  mode-frequencies
corresponding to the second case
[simply obtained by solving the quadratic (\ref{disper})] for the particular
value $x_\p=0.05$. This figure shows the presence of unstable
modes within the range of values for $\eps$ that we take to 
be physically realistic \citep{prix}: $0.4\le \eps \le 0.7$. 
From this figure we immediately deduce two things. First, we see that the 
unstable modes indeed satisfy the criterion (\ref{yrange}). 
Secondly, the unstable modes may have imaginary parts as 
large as Im~$\kappa \approx 0.15$. It is useful to 
ask what this implies for the growth time of the instability.
Since our results only depend on the azimuthal index $m$ 
through the scaling 
\begin{equation}
\mbox{Im } \omega = m \Omega_\p \mbox{ Im }\kappa
\end{equation}
we see that the fastest growth time corresponds to the $m=l$ modes.
The e-folding time for the instability follows from
\begin{equation}
t_e =  { 1 \over m \Omega_\p \mbox{ Im }\kappa }= 
{ 1 \over 2 \pi m \mbox{ Im }\kappa}  P \mbox{ s}
\end{equation} 
where $P$ represents the observed rotation period of the pulsar
(presumably corresponding to the rotation of the charged component, i.e.
$P/2\pi/\Omega_\p$). 

The above example shows that the two-stream instability 
does indeed operate in this shell problem. 
In fact, the analysis goes beyond the local analysis 
of the plane parallel problem in Section~IIB since we have 
now solved for the actual unstable modes (satisfying the 
relevant boundary conditions). We see that, as expected for a dynamical instability,
the growth time of an unstable mode may be very short. 
However, the relative rotation rates required to make 
the quadrupole modes unstable in the range $0<x_\p<1$ and
$0<\eps<1$ are likely far too large to be physically
attainable. In this sense the results shown in Figs.~\ref{shell}--\ref{rotfreq}
are, despite being instructive, somewhat extreme. 

\subsection{A realistic mechanism for triggering pulsar glitches?}

A quantity of key importance for this discussion is the
rotational lag between the two components. In order to be able to argue 
that the two-stream instability is relevant for pulsar  glitches
we need to 
consider lags that may actually occur in astrophysical 
neutron stars. To estimate the size of the rotational lag required
to ``explain'' the observed glitches we assume
 that a glitch  corresponds to a transfer of 
angular momentum from a partially decoupled superfluid component 
(index $\n$) to the bulk of the star (index $\p$). Then we have
\begin{equation}
I_\n |\Delta \Omega_\n| \approx I_\p \Delta \Omega_\p
\rightarrow \Delta \Omega_\p \approx { I_\n \over I_\p} |\Delta \Omega_\n|
\end{equation}
where $I_X$ are the two moments of inertia. Now assume that the 
decoupled component corresponds 1\% of the total moment of inertia, eg. 
the superfluid neutrons in the 
inner crust or a corresponding amount of fluid in the core. 
This would mean that $I_\n \sim 10^{-2} I_\p$, and we have
\begin{equation}
\Delta \Omega_\p \approx 10^{-2} |\Delta \Omega_\n| \ . 
\end{equation}
Combine this with the 
observations of large Vela glitches to get
\begin{equation}
{ \Delta \Omega_\p \over \Omega_\p} \approx 10^{-2} { |\Delta \Omega_\n|
\over \Omega_\p} \sim 10^{-6} \ . 
\end{equation}
In other words, we must have
\begin{equation}
|\Delta \Omega_\n| \approx 10^{-4} \Omega_\p \ . 
\end{equation}
If we assume that the glitch brings the two fluids back into 
co-rotation, then we have $\Delta \Omega_\n = \Omega_\n-\Omega_\p = \Delta \Omega$
and we see that the two rotation rates will 
maximally differ by one part in $10^4$ or so. Rotational lags of this 
order of magnitude have often been discussed in the context of glitches. 
Even though the key quantity in models invoking catastrophic
vortex unpinning in the inner crust --- the pinning strength --- is very uncertain, 
and there have been suggestions that the pinning force is too weak 
to allow a build up of the required rotational lag \citep{jones}, typical values 
considered are consistent with our rough estimate.
In addition, frictional heating due to a difference in the 
rotation rates of the bulk of a neutron star and a superfluid component 
has been discussed as a possible explanation for the fact that 
old isolated pulsars seem to be somewhat hotter than expected from standard
cooling models \citep{friction1,friction2}.  \citet{friction2}
argue that a lag of
\begin{equation}
{\Delta \Omega \over \Omega_\p}  \approx (3.2\times10^{-4}-9.5\times 10^{-3})\times 
\left(0.1\mbox{ s} \over P \right)
\end{equation}
could explain the observational data.
Finally, the presence of rotational lags of the proposed magnitude is supported
by a statistical analysis of 48 glitches in 18 pulsars \citep{lyne}. This study 
suggests that the critical rotational lag at which a glitch occurs is
\begin{equation}
{ \Delta \Omega \over \Omega_\p} \approx 5 \times 10^{-4} \ . 
\end{equation}

\begin{figure} 
\centering
\includegraphics[height=8cm,clip]{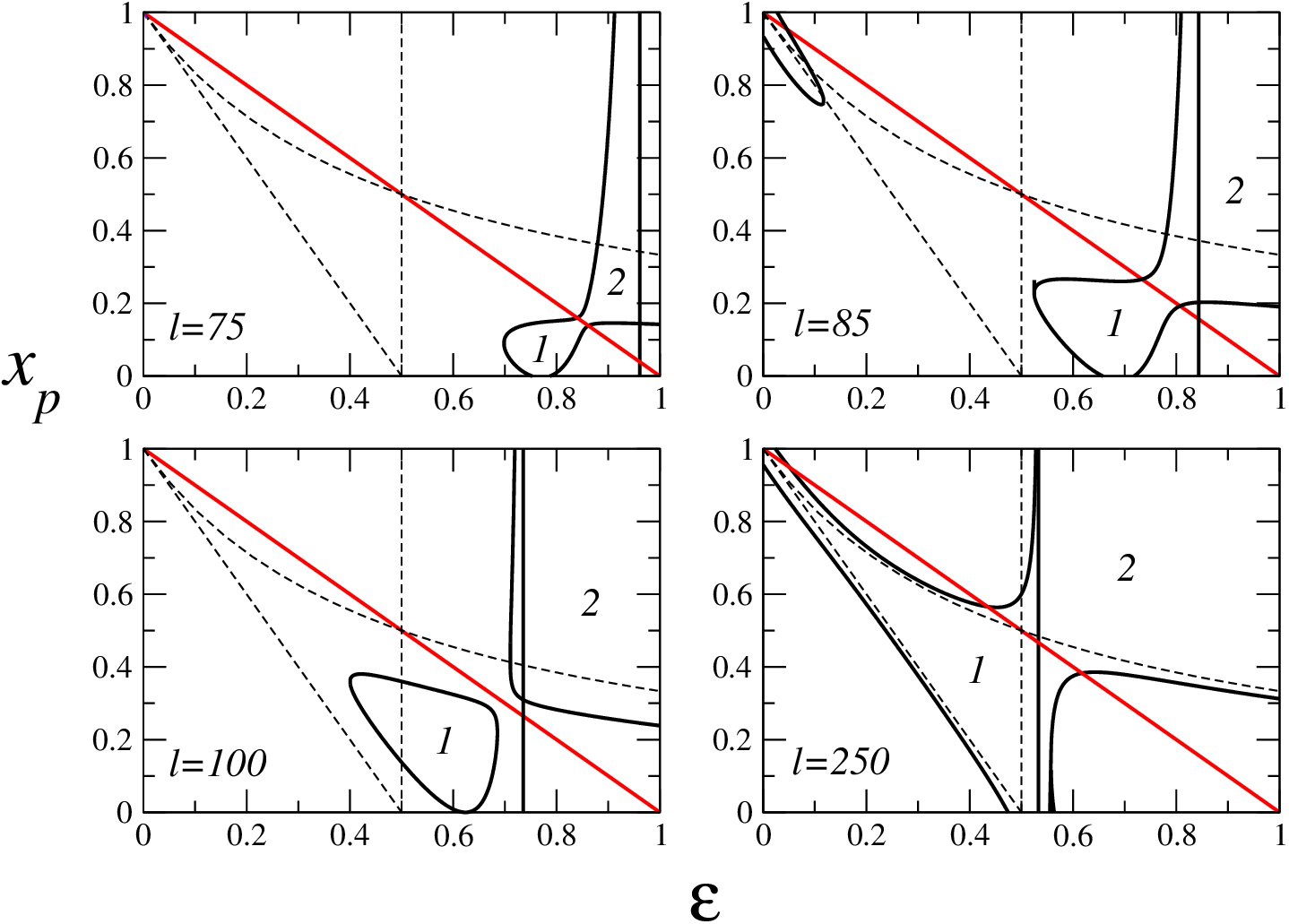}
\caption{ The two-stream instability regions (labeled 1 and 2) 
for the case when the 
superfluid neutrons lag behind the superconducting protons
in such a way that $\Delta \Omega /\Omega_\p 
= 5\times 10^{-4}$ or $\y = 1.0005$. We show results for 
four different values of $l$. There are no unstable 
modes in this part of parameter space unless $l>65$. Recall that 
$x_\p<0.1$ and $0.4\le \eps \le 0.7$ would be reasonable parameter
values for a neutron star core. The grey diagonal line represents the 
singularity discussed in Section~IIID and the dashed curves indicate the 
instability regions in the large $l$ limit.}
 \label{rotdomain}
\end{figure}

\begin{figure} 
\centering
\includegraphics[height=8cm,clip]{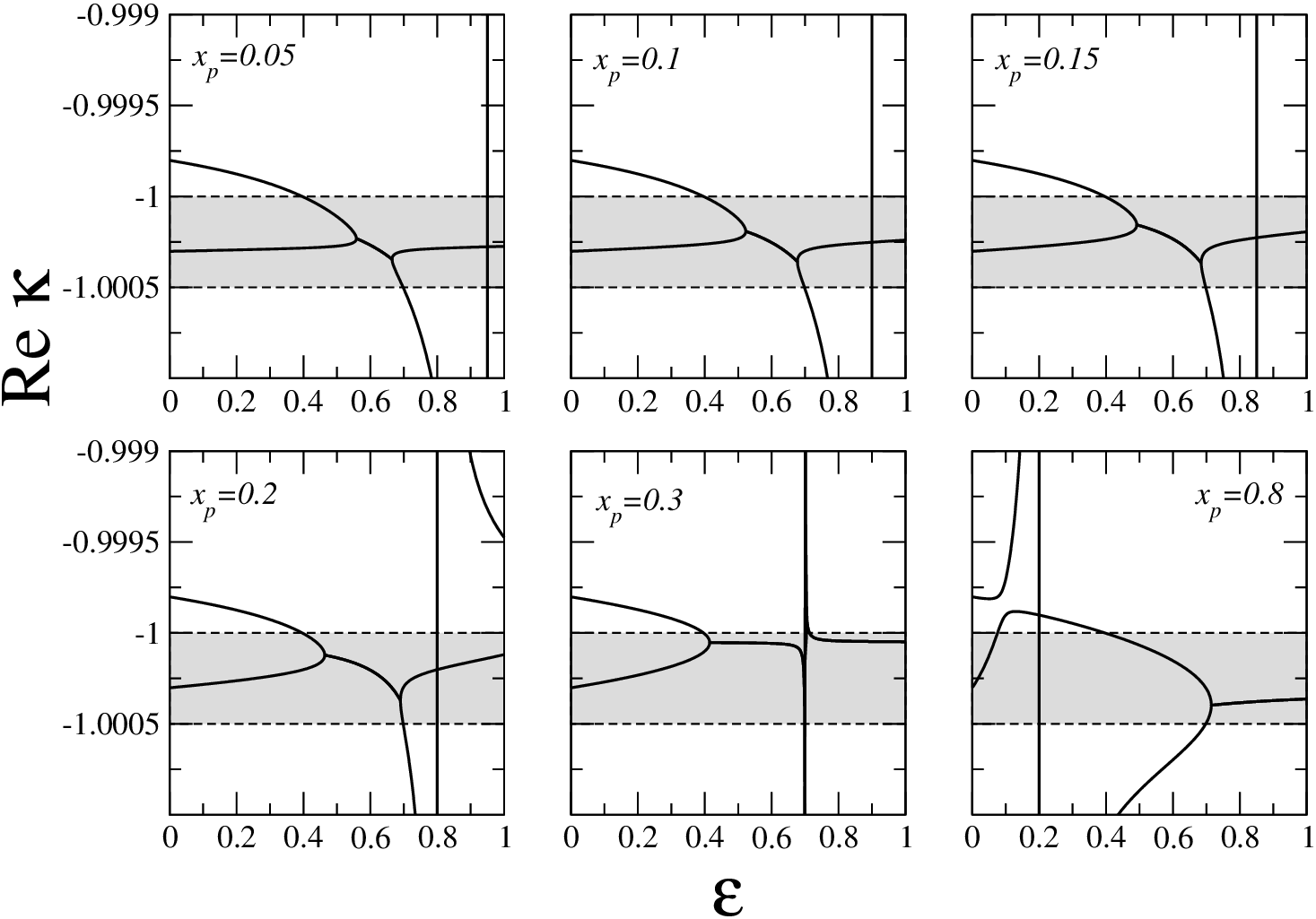}
\caption{Mode frequencies (represented by 
Re~$\kappa$) as function of the entrainment parameter $\eps$ 
for $l=100$,  $\y=1.0005$ and various 
values of $x_\p$. The range in which an instability is permissible
[according to (\ref{yrange})] 
is indicated by the grey areas.}
\label{freq25}
\end{figure}

\begin{figure} 
\centering
\includegraphics[height=6cm,clip]{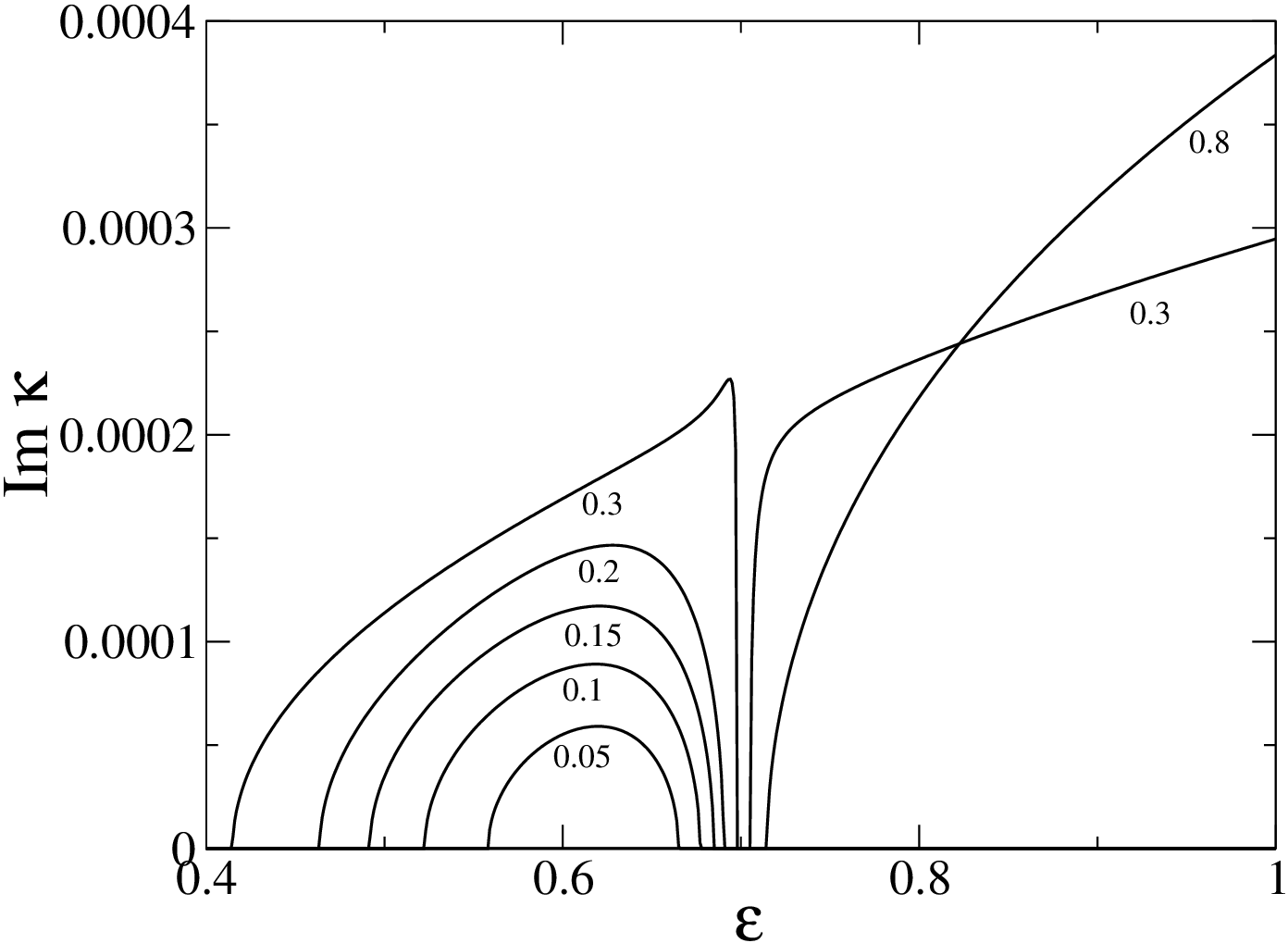}
\caption{The imaginary part Im~$\kappa$ (determining the growth rate) 
of the unstable modes for  
$\y=1.0005$ and $l=100$ is shown as a function of $\eps$ 
for several different values of $x_\p$ (as indicated in the figure). 
Essentially, these results represent 
horizontal cuts through the frames in Figure~\ref{rotdomain}.  }
\label{imag25}
\end{figure}

In order to make our shell model problem more realistic we 
 consider the  case when the superfluid neutrons
lag behind the superconducting protons,  and  take 
$\Delta \Omega /\Omega_\p = 5\times 10^{-4}$, or $\y = 1.0005$, as a
 representative value. With this rotational lag, we see from
 (\ref{yrange}) that the unstable modes must be such that 
$-1.0005 < \kappa < -1$.    
A series of results for this choice of parameters 
are shown in Figures~\ref{rotdomain}-\ref{imag25}. 

Figure~\ref{rotdomain} illustrates the fact that, if we 
decrease the rotational lag then the two-stream instability 
will not be active (in the interesting region of
parameter space) for low values of $l$.  
For a smaller rotational lag the instability 
acts on a shorter length scale. For $\y = 1.0005$ we find that 
we must have $l>65$ in order for there to be a region of instability
in the part of the $x_\p-\eps$ plane shown in 
Figure~\ref{rotdomain}. This means that the instability 
only operates on length scales shorter than $\pi R/l \approx 500 $~m
(if we take the shell radius to be $R=10$~km). 

Figure~\ref{freq25} shows the real part of the mode frequencies 
for $l=100$ and various values of $x_\p$. From this figure we can see that 
the instability always occur in the region suggested by the 
instability condition (\ref{yrange}), i.e. two real frequency modes never merge 
to give rise to a complex conjugate pair of solutions outside the grey areas 
indicated in  the various panels of Figure~\ref{freq25}.
The imaginary parts for $l=100$ and the various values of $x_\p$
considered in Figure~\ref{freq25}
are shown in Figure~\ref{imag25}.
From this figure we see that the imaginary part of $\kappa$ typically reaches 
values of order $10^{-4}$. In fact, 
by comparing similar results for different values of $l$ 
we have found that the largest attainable imaginary part of $\kappa$ varies
by less than one order of magnitude as $l$ increases from
100 to 1000. Thus we estimate that the typical instability growth time 
for $\y=1.0005$  will be 
\begin{equation}
t_e \approx { 10^4 \over m } { P \over 2\pi} \mbox{ s} \ . 
\end{equation}
For a star rotating at the rate of the Vela pulsar, $P= 89$~ms, 
we would have  $t_e \approx 1.4$~s for $l=m=100$.
Interestingly,  this predicted growth time is significantly
shorter than the resolved rise time of a large Vela glitch
$t_{\rm glitch}< 40$~s \citep{velarise}.

\subsection{Approximate results: The large $l$ limit}

The results obtained above indicate that the two-stream instability
is likely to act on modes with rather short wavelengths. Given this
it makes sense to consider the large $l$ limit 
in more detail. Keeping only the leading order term (proportional to $l^4$)
in (\ref{disper})
we have the dispersion relation
\begin{eqnarray}
(1-\eps-x_\p)\kappa^2  & +&
\left[ 1-2\eps-x_\p+2x_\p \eps + (1-x_\p-2x_\p \eps)\y\right]\kappa \nonumber \\
&+&
(1-2\eps-x_\p)\y +\eps x_\p + \eps(1-x_\p)\y^2-2x_\p\eps^2(1-\y)^2 =0 \ . 
\end{eqnarray} 
We can readily write down the solutions to this quadratic:
\begin{eqnarray}
\kappa_\pm &=& { 1 \over 2(1-x_\p-\eps)} \left\{ 1-2\eps-x_\p+2\eps x_\p 
+(1-x_\p-2x_\p \eps) \y \right. \nonumber \\
& \pm& \left. (1-\y) \left[(1-2\eps)(1-2\eps-x_\p)(1-x_\p-2\eps x_\p) \right]^{1/2}
\right\} \ . 
\label{kappas}\end{eqnarray}
It turns out that we can learn a lot about the problem from this
expression. The most obvious feature is the fact that 
$\kappa_\pm$ have a singularity when $1-x_\p-\eps=0$.
It is straightforward to show that one of the roots will become infinite
at this point, while the second root becomes:
\begin{equation}
\kappa = { 1 \over 1-2 x_\p} \left[ \y - x_\p(1+\y)+2x_\p \eps (1-\y) \right] \ . 
\end{equation} 
For the examples shown in Figure~\ref{rotdomain}
this special case corresponds to the intersection between 
the instability domain and the 
diagonal line $1-x_\p-\eps=0$, as  indicated in 
Figure~\ref{rotdomain}. 

It is also straightforward to deduce the regions of instability from 
the sign of the argument of the square-root. We see that we will have 
an unstable mode when
\begin{equation}
\eps<{ 1\over 2} \qquad \mbox{and} \qquad 1-2\eps<x_\p<{ 1 \over 1+2\eps}
\qquad \mbox{Region 1}\end{equation}
or
\begin{equation}
\eps>{ 1\over 2} \qquad \mbox{and} \qquad x_\p>{ 1 \over 1+2\eps}
\qquad \mbox{Region 2} \ . 
\end{equation}
These regions are also indicated in Figure~\ref{rotdomain}.
Out of these two possible instability regions, the first is  most likely to 
be relevant for neutron stars since it allows the instability to be present
already for small proton fractions. 

Finally, we can  use (\ref{kappas}) to show the existence of an extremum at
\begin{displaymath}
\eps = { 1 \over 8} ( \sqrt{17}-1)\approx 0.39 \quad \ , \quad x_\p = 
{ 1\over 4} {\sqrt{17}+7 \over \sqrt{17}+3} \approx 0.39 \ . 
\end{displaymath}
The corresponding imaginary part of $\kappa$ would represent
the fastest possible growth time for an unstable mode located in
region~1 of Figure~\ref{rotdomain}. This leads to the estimate
\begin{equation}
\mbox{Im }\kappa \le 0.24 (\y-1)  
\end{equation}
or an estimate of the shortest growth time:
\begin{equation}
t \approx { 6.7\times 10^{-2}\over l} \left( { \Delta \Omega \over \Omega_\p} \right)^{-1} 
\left({ P \over 0.1 \mbox{ s}} \right) \mbox{ s}
\label{growth}\end{equation}
which agrees well with the result we previously obtained for $l=m=100$.
Hence, this simple formula can be used to estimate  the 
fastest growth rate of the two-stream instability in our shell model
for different parameter values.

\section{Discussion}

In this paper we have introduced the superfluid two-stream instability:
A dynamical instability analogous to that known
to operate in plasmas \citep{anderson}, which sets in once the relative 
flow between the two components of the system reaches a critical level. 
We have studied this instability for two different model problems. 
First we analysed a local dispersion relation derived for the case of a background
such that one fluid was at rest while the other had a constant flow rate.
This provided a proof of principle of the existence of the two-stream 
instability for superfluids. Our analysis was based on the two-fluid equations
that have been used to model the dynamics of the outer core of a neutron 
star, where superfluid neutrons are expected to coexist with superconducting
protons and relativistic electrons. These equations are analogous to the 
Landau model for superfluid Helium~\footnote{Even though we do not discuss this 
issue in detail here, it is exciting to contemplate possible experimental 
verification of the superfluid two-stream instability in, for example, 
superfluid $^4$He.}, and should also (after some modifications
to incorporate elasticity and possible vortex pinning) be relevant 
for the conditions in the inner crust of a mature neutron star.
Thus we expect the two-stream instability to be generic in dynamical
superfluids, possibly limiting the relative flow rates of 
any multi-fluid system. Our second model problem concerned two fluids
confined within an infinitesimally thin spherical shell.
The aim of this model was to assess whether 
the two-stream instability may be relevant (perhaps as trigger mechanism)
for pulsar glitches. The results for this problem demonstrated 
that the entrainment 
effect could provide a sufficiently strong coupling for the instability
to set in at a relative flow small enough to be astrophysically 
plausible. Incidentally, the modes  
that become dynamically unstable in this problem are the superfluid analogues of the 
inertial r-modes of a rotating single fluid star. This is interesting 
since the r-modes are known to be secularly unstable due to the emission
of gravitational radiation~\citep{r_rev}. In fact, the connection between the two 
instabilities goes even deeper than this since the radiation-driven secular
instability is also a variation of the Kelvin-Helmholtz instability.
In that case, the two fluids are the stellar fluid and the radiation.

In order for an instability to be relevant
the unstable mode must grow faster than all  
dissipation timescales. In the case of a superfluid 
neutron star core the main dissipation mechanisms are 
likely to be mutual friction and ``standard'' shear viscosity
due to electron-electron scattering. Since we have tried to 
build a plausible case for the two-stream instability to
be relevant for pulsar glitches we would like to obtain some
rough estimates of the associated dissipation timescales. 
To do this we first use 
an estimate of when mutual friction is likely to dominate
the shear viscosity [due to  \citet{mendell2}]: 
\begin{equation}
\Omega > 100 \left( {10^6 \mbox{ cm} \over \lambda}\right)^2 \left({T \over 10^7\mbox{ K}} \right)^2 
\mbox{s}^{-1} \approx 100 \left( {l \over \pi}\right)^2 \left({T \over 10^7\mbox{ K}} \right)^2 
\mbox{s}^{-1}
\end{equation}
where we assume that the wavelength of the mode is $\lambda = \pi R /l$. We can write this as
\begin{equation}
P < 0.62 l^{-2}\left({T \over 10^7\mbox{ K}} \right)^2   \mbox{ s}
\end{equation}
which suggests that the shear viscosity will be the dominant 
dissipation mechanism for large values of $l$. 
For example, for a neutron star rotating with the period of the Vela pulsar
mutual friction would dominate for $l<15$ or so (assuming $T_7\approx 5$).
To estimate the shear viscosity damping we can use results obtained for the 
secular r-mode instability. In particular,
\citet{ksteg} have shown that for a uniform density star
one has
\begin{equation}
t_{sv} \approx { 3 \over 4\pi(l-1)(2l+3) } { M \over \eta R} \approx 
{ 3 \over 8\pi l^2}  { M \over \eta R} \ . 
\end{equation}
If we use the shear viscosity coefficient for electron-electron scattering
\begin{equation}
\eta_{\e\e} \approx 6 \times10^{18} \left( {\rho \over 10^{15} \mbox{ g/cm}^3} \right)^2 
\left({ 10^9\mbox{ K} \over T} \right)^2  \mbox{ g/cm s}
\approx 2.7\times 10^{18} \left( { M\over 1.4 M_\odot} \right)^2 
\left({10 \mbox{ km} \over R}\right)^{6}\left({ 10^9\mbox{ K} \over T} \right)^2   \mbox{ g/cm s}
\end{equation}
we get
\begin{equation}
t_{sv} \approx  { 1.2\times10^4 \over l^2 } \left( { 1.4 M_\odot} \over M \right) 
\left({R \over 10 \mbox{ km} }\right)^{5}\left({T \over 10^7\mbox{ K}} \right)^2  \mbox{ s}
\label{tsv}\end{equation}

We want to compare this damping timescale to the growth rate of the unstable modes
in our shell toy-model. Combining (\ref{growth}) with (\ref{tsv}) we estimate that in order
to have an instability we must have
\begin{equation}
l < 90 \left( {\Delta \Omega / \Omega_\p \over 5\times10^{-4}}\right) 
\left({ 0.1 \mbox{ s} \over P} \right)
\left( { 1.4 M_\odot} \over M \right) 
\left({R \over 10 \mbox{ km} }\right)^{5} \left({T \over 10^7\mbox{ K}} \right)^2 \ . 
\end{equation}
Let us now consider the case of the Vela pulsar. 
Estimating the core temperature as $5\times 10^7$~K [roughly two orders of
magnitude higher than the observed surface temperature \citep{page}] we deduce that only modes 
with $l>2500 $ or so are likely to be stabilized by shear viscosity.  
Given that our results indicate that the two-stream instability 
is active for much smaller values of $l$, cf. the results shown in Figure~\ref{rotdomain}, 
we conclude the dissipation is unlikely to suppress the
instability in sufficiently young neutron stars. 
Incidentally, the length scale corresponding to a mode with $l=2500$ 
would be about ten meters.
This is an interesting result since one can show that a 
large glitch could be explained by a small fraction ($\sim 10^{-4}$)
of the neutron vortices moving a few tens of metres \citep{cordes}.

Obviously, the situation changes as
the star cools further. Based on the above estimates one can show that 
shear viscosity will suppress all modes with $l>65$ (i.e. all unstable modes
for the case considered in Fig.~\ref{rotdomain}) if 
the core temperature is below $8\times 10^6$~K.
This means that the two-stream instability may not be able to overcome the 
viscous damping in a sufficiently cold neutron star, which is
consistent with the absence of glitches in mature pulsars.  

We believe that
the results of this paper suggest that the superfluid two-stream 
instability may be relevant in the context of pulsar 
glitches. If this is, indeed, the case then what is its exact role?
The answer to this question obviously requires much further work, but 
it is nevertheless interesting to speculate about some possibilities.
Most standard models for glitches are based on the idea of catastrophic
vortex unpinning in the inner crust \citep{ai}. This is an attractive 
idea since the glitch relaxation (on a timescale of days to months) would
seem to be well described by vortex creep models \citep{vcreep}.
An interesting scenario is provided by the thermally induced glitch model
 discussed by \citet{link1}. They have
shown that the deposit of $10^{42}$~erg of heat would be sufficient
to induce a Vela type glitch. 
The mechanism that leads to the unpinning of vortices, eg. by the deposit
of heat in the crust, is however not identified. We 
believe that the two-stream instability may fill this gap 
in the current theory. 
It should, of course, be pointed out that glitches need not originate in 
the inner crust. In particular,  \citet{jones}
has argued that the vortex pinning 
is too weak to explain the recurrent Vela glitches. If this argument 
is correct then the glitches must be due to some mechanism operating in the 
core fluid. Since the model problems we have considered would be relevant for 
the conditions expected to prevail in the outer core of a mature neutron star, 
our results show that the two-stream instability may serve as a trigger
for glitches originating there. The key requirement for the 
instability to operate is the presence of a rotational lag. 
It is worth pointing out that such a lag will build up both when there
is a strong coupling between the two fluids (i.e. when the vortices are pinned)
and when this coupling is weak. One would generally expect the strength 
of this coupling to vary considerably at various depths in the star \citep{langlois}, and it 
is not yet clear to what extent a rotational lag can build up 
in various regions. This is, of course, a key issue for  future
theoretical work on pulsar glitches.

One final relevant point concerns the recent observation of a 
Vela size glitch in the anomalous X-ray pulsar 
1RXS J170849.0-400910 \citep{kaspi}. This object 
has a spin period of 11~s, which means that any 
feasible glitch model must not rely on the star 
being rapidly rotating. What does this mean for our proposal 
that the two-stream instability may induce a glitch?
Let us assume that the  rotational lag builds up at the same rate as the 
electromagnetic spindown of the main part of the star (i.e. that 
the superfluid component does not change its spin rate at all
under normal circumstances).  
Then the lag would be $\Delta \Omega \approx t \dot{\Omega}$ after time $t$. 
If there is a critical value at which a glitch will happen (corresponding to
$\Delta \Omega_{\rm crit}$) then the interglitch time $t_g$ could be
approximated by
$$
t_g \approx { \Delta \Omega_{\rm crit} \over \Omega } { \Omega \over \dot{\Omega}} = 
2\tau { \Delta \Omega_{\rm crit} \over \Omega } 
$$  
where $\tau$ is the standard ``pulsar age''. This argument implies the following:
i) For $ \Delta \Omega_{\rm crit} / \Omega\approx 5\times10^{-4}$
we would get $t_g \approx 10^{-3} \tau$. This (roughly) means that only pulsars younger than
 $10^4$~yr would be seen to glitch during 30 years of observation, 
which accords well with the fact that only young pulsars are 
active in this sense. ii) There is no restriction on the rotation rate in this 
scenario; a star spinning slowly may well exhibit a glitch as long as  
its spindown rate is fast enough. This means that one should not 
be surprised to find glitches in stars with 
extreme magnetic fields (magnetars). 

This paper is only a first probe into what promises to 
be a rich problem area. Future studies must address issues concerning
the effects of different dissipation mechanisms, the nonlinear evolution
of the instability, possible experimental verification for superfluid 
Helium etcetera. These are all very interesting problems which we 
hope to investigate in the near future.

\section*{Acknowledgements}

NA and RP acknowledge support from the EU
Programme 'Improving the Human Research Potential and the Socio-Economic
Knowledge Base' (Research Training Network Contract HPRN-CT-2000-00137).
NA acknowledges support from the Leverhulme Trust in the form of a prize 
fellowship, as well as generous hospitality offered by the Center for 
Gravitational-Wave Phenomenology at Penn State University. 
GC acknowledges partial support from NSF grant PHY-0140138.

\end{document}